\documentclass[preprint,preprintnumbers,amsmath,amssymb,showpacs]{revtex4}
\usepackage{graphicx}
\usepackage{dcolumn}
\usepackage{bm}

\begin{document}

\title{Interference effects in the Coulomb dissociation of $^{15,17,19}$C.}
\author{T. Tarutina}
\email{tatiana@fma.if.usp.br}
\author{M.S. Hussein}
\affiliation{Departamento de F\'isica Matem\'atica, Instituto de F\'isica
  da Universidade de S\~ao Paulo, \\
Caixa Postal 66318, 05315-970 S\~ao Paulo, SP, Brazil}

\date{\today}

\begin{abstract}
In this work the semiclassical model of pure Coulomb excitation
was applied to the  breakup of $^{15,17,19}$C. The ground state
wave functions were calculated in the particle-rotor model
including core excitation. The importance of interference terms in
the dipole strength arising after including core degrees of
freedom is analyzed for each isotope. It is shown that Coulomb
interference effects are important for the case of $^{17}$C.

\end{abstract}


\maketitle

\section{Introduction}
The study of neutron rich light exotic nuclei has been the subject
of extensive experimental and theoretical work for more than two decades already.
One of the tools to probe the structure of these nuclei are the
experiments on the Coulomb excitation and breakup whose reaction
mechanism is very well known.

Several experiments have been recently performed  to study carbon
isotopes $^{15,17,19}$C. It was found \cite{Baz95}  that $^{19}$C is a candidate
to have one-neutron halo. Ground state with spin-parity
$\frac{1}{2}^+$ \cite{Baz95,Nak99,Mad01} and one-neutron separation energy less than 1 MeV
\cite{Baz95,Nak99,Mad01}
favors the formation of the halo in this nucleus, which is also
confirmed by narrow momentum distributions of $^{18}$C following
$^{19}$C breakup \cite{Baz95}.

 There were
several experiments to measure separation energy of $^{19}$C.
The experiments using time-of-flight techniques suggest small
 separation energy, that is, weighted average yields 
 242$\pm$95 keV \cite{Baz95}.
 The Coulomb dissociation of $^{19}$C was studied by Nakamura
in \cite{Nak99}. The analysis of angular distributions of breakup
products suggests the value 0.53$\pm$0.13 MeV. Using this value in
the simple  cluster model calculation of the dipole strength gives
good agreement with the data.  But the
analysis of recent experiment of Maddalena {\it et al.}
\cite{Mad01} on nuclear breakup of $^{19}$C yields 0.65$\pm$0.15
MeV and 0.8$\pm$0.3 MeV. The adopted value of one-neutron separation
energy of $^{19}$C is given in Ref. \cite{Aud03} and $S_n=0.58$ MeV.

Unlike $^{19}$C, the spin-parity of $^{17}$C was found to be
$\frac{3}{2}^+$ \cite{Mad01}, with binding energy also smaller 
then 1 MeV \cite{Fif82}.
However the spin $\frac{3}{2}^+$ is an indication of a $d$-wave single
particle structure and accordingly the existence of centrifugal
barrier in this case does not favor the halo.

$^{15}$C again has a $\frac{1}{2}^+$ ground state \cite{Gos73} with a large
amount of $s$-wave in the wave function with the separation energy
of around 1.2 MeV \cite{Aud03}. Although this isotope is not generally
recognized as a halo-nucleus, there is an evidence that this
nucleus has a halo-like structure (see \cite{Par00} and
\cite{Cha00}).

Coulomb breakup of $^{15}$C
and $^{17}$C at relativistic energies has recently been studied 
in Ref. \cite{Pra03}. The main ground state configuration of $^{15}$C
is found to be $^{14}$C$(0^+)\otimes\nu_{s}$ with spectroscopic 
factor consistent with earlier studies. The predominant ground state
configuration is of $^{17}$C is found to be $^{16}$C$(2^+)\otimes\nu_{s,d}$.

A number of theoretical studies of Coulomb and nuclear breakup of
heavy carbon isotopes were performed in the framework of post-form
distorted wave Born approximation \cite{Cha00,Shy01,Ban02,Cha03}.
The study of Ref. \cite{Cha00} confirmed the existence of
one-neutron halo in $^{11}$Be,$^{19}$C and $^{15}$C but not in
$^{17}$C. In Ref. \cite{Shy01} Coulomb breakup of $^{11}$Be and
$^{19}$C on $^{208}$Pb was analyzed, where the transitions to
excited states of the projectile´s core were calculated. It was
found that the contributions of the core excited states are very
small. In Ref. \cite{Ban02} the respective roles of the first order
and higher order effects for different beam energies were
investigated. It was  found that higher order effects are small in
case of higher beam energies and forward scattering, but important
at incident energies $\le$ 30 MeV/nucleon. In Ref. \cite{Typ01} a
semiclassical dynamical model of projectile excitation was applied
to nuclear and Coulomb breakup of $^{11}$Be and $^{19}$C on
$^{208}$Pb. It was shown that Coulomb breakup dominates relative
energy spectra around the peak region while the nuclear breakup is
important at higher relative energy. The higher order effects are
found to be generally small and dependent on the theoretical
model.

In this paper we use the semiclassical model of pure Coulomb breakup.
In this model the Coulomb breakup cross-section is calculated
using the virtual photon numbers formalism and the dipole strength
function $dB(E1)/dE$ which requires the knowledge of the ground
state and excited states wave functions. For the ground state we
use wave functions obtained in the particle-rotor model with core
excitation. For the continuum states we use plane waves. The
inclusion of core degrees of freedom gives rise to the
interference terms in the dipole strength function. The model
employed here was used before in Ref. \cite{Rid98}, where the
relative energy spectra were calculated for the Coulomb breakup of
$^{19}$C and integrated cross-sections for the Coulomb breakup of
$^{15,17,19}$C were presented for different ground state scenarios
of these carbon isotopes. The lack of experimental data made it
difficult to draw final conclusions about structure of these heavy
carbon isotopes.

The purpose of this paper is to give a detailed analysis of the
particle-core result.
 In particular we assess the importance of
the interference terms in the dipole strength function, linear in
the core deformation, on the shape and value of the cross section. In our
approach we shall analyze the nuclear-corrected data, namely the
ones where the nuclear contribution has been removed
 through the commonly used extrapolation procedure. For a recent overview 
of this scaling method see Ref. \cite{Nak04}.

\section{Core-particle model with core excitation}
The inclusion of core degrees of freedom as it is done in the
core-particle model with core excitation is simple and physically
transparent. This method was described in \cite{Boh75}.
It was employed in \cite{Nun95,Nun961,Nun962} to study the properties of such weakly-bound and
unbound two-body systems as  $^{11}$Be, $^{13}$C, $^{10}$Li and
three-body $^{12}$Be. The
inclusion of core excitation made it possible to describe
$^{11}$Be as an $s$-intruder nucleus. Esbensen {\it et
al.}~\cite{Esb95} also applied this model to positive parity states
in $^{11}$Be and $^{13}$C.
In the
work by Ridikas {\it et al.} \cite{Rid98} different scenarios
were proposed for the g.s. structure of $^{15,17,19}$C.
 Recently this model was applied to the
simultaneous description of borromean nucleus $^{14}$Be and its
binary constituent $^{13}$Be \cite{Tar04}.
$^{11}$Be was also studied in the particle-vibrator model by Vinh Mau
in \cite{Mau95}.

The basic idea of this method is that deformation of the core can lead
to couplings with excited states of the core.
The total Hamiltonian $\hat H$ of the two-body system(${core + n}$)
can be written in the following way:

\begin{equation}
\hat H = \hat T + \hat H_{rot} + \hat V,
\end{equation}
where $\hat T$ is kinetic energy of the relative motion of core
and valence neutron, $\hat H_{rot}$ is the Hamiltonian of a
deformed axially symmetric rigid rotor, $\hat V$ is the
interaction between the core and the neutron.  The wave function
of the system with total spin $J$ has the following form:
\begin{eqnarray}
\nonumber
 \Psi^{JM} = \sum_{ljI}\sum_{m_lm_sm_jm_I}
\frac{\chi_{ljI}^{J}(r)}{r} \langle lm_lsm_s|jm_j\rangle\times
\\\langle jm_jIm_I|JM\rangle
 Y_{lm_l}(\hat r)X_{sm_s}(\hat\sigma)
\phi^{I}_{m_I0}(\hat\xi), \label{wf}
\end{eqnarray}
where $\chi_{ljI}^{J}(r)$ are the radial wave functions,
$Y_{lm_l}(\hat r)$ are the spherical
harmonics, $X_{sm_s}(\hat\sigma)$ are the spin functions.
 The core
states $\phi^{I}_{m_I0}(\hat\xi)$ are eigenvalues of the $\hat
H_{rot}$ and are proportional to the rotational matrices:
$$
\phi^{I}_{m_I0}(\hat\xi)=\frac{\hat I}{\sqrt{8\pi^2}}{\mathcal
D}^I_{m_I0}(\hat\xi),
$$
where $\hat\xi$ are the Euler angles, characterizing the
orientation of the core in the laboratory system and $\hat
I\equiv\sqrt{(2I+1)}$.

 The interaction term
in the Hamiltonian consists of the deformed Woods-Saxon potential
and  standard undeformed spin-orbit part:
\begin{eqnarray}
\nonumber \hat V = V^{ws}(r,\theta^{'})+{\bf l} \cdot{\bf
s}V^{so}(r)\\
V^{ws}(r,\theta^{'})= \frac{V_0^{ws}}{1+\exp\frac{r-R(\theta')}{a}}\\
V^{so}(r)=2(\frac{\hbar^2} {m_{\pi}c})^2
\frac{V_{0}^{so}}{r}\frac{d}{dr}\left\{1+\exp\frac{r-R_0}{a}\right\}^{-1}.
\end{eqnarray}

We include only quadrupole term  in the expansion of the radius:
\begin{equation}
R(\theta^{'}) = R_0\left[1+\beta Y_{20}(\cos\theta^{'})\right],
\end{equation}
where $\beta$ is the quadrupole deformation parameter and
$\theta'$ is the polar angle of the particle in the body-fixed
coordinate system. 

 To solve the problem we substitute the expansion (2) into
the total Schr\"odinger equation to obtain the set of coupled
equations:
\begin{equation}
\left(T + V_{\gamma\gamma}^{J}(r)
-E+\epsilon_I\right)\chi_{\gamma}^{J}(r) =
-\sum_{\gamma^{'}\ne\gamma}V_{\gamma\gamma^{'}}^{J}(r)\chi_{\gamma^{'}}^{J}(r),
\label{ccs_main}
\end{equation}
where $\gamma=\{l,j,I\}$,
kinetic energy
$T =-\hbar^2/2\mu\left[d^2/dr^2 -
  l(l+1)/r^2\right]$
and matrix elements $V_{\gamma\gamma'}(r)= \langle l,j,I;JM |\hat V(r,\theta')|l',j',I';JM\rangle$.
$\epsilon_I$ is the excitation energy of the core state with spin $I$.

The deformed part of the interaction is then expanded in terms of the
Legendre polynomials in order to separate the angular and radial part
in the calculation of the matrix elements:
\begin{equation}
V(r,\theta')=\sum_Q V_Q(r)P_Q(\cos\theta')
\end{equation}
The corresponding angular part of the matrix elements could be found
in \cite{Boh75}.

The alternative and further simplified way to calculate matrix
elements is to expand the deformed Woods-Saxon potential in Taylor
series over deformation parameter $\beta$ keeping only the leading
term:
\begin{eqnarray}
V(r,\theta')=V_0f(r)+ \beta v(r)Y_{20}(\cos\theta')\\
f(r)=\frac{1}{1+\exp{\frac{r-R_0}{a}}} \label{taylor}
\end{eqnarray}
where $v(r)=V_0f(r)^2\exp\left(\frac{r-R_0}{a}\right)\frac{R_0}{a}$
is the radial part of the coupling matrix element.
This method allows the analytical calculation of the radial part of
the coupling matrix elements.

In our analysis we will employ particle-core method with core
excitation to describe $^{19}$C and $^{17}$C. We also use this
model to describe $^{15}$C, although the core $^{14}$C is not a rotor.
It is already established that the ground state of the $^{19}$C is
1/2$^+$ state.  Thus, for the ground state of $^{19}$C and $^{15}$C
we will have to couple following three channels:
$$
|\frac{1}{2}^+\rangle=a_1|2s_{1/2}\otimes 0^+\rangle + a_2|1d_{5/2}\otimes
 2^+\rangle + a_3|1d_{3/2}\otimes 2^+\rangle,
$$
For $\frac{3}{2}^+$ ground state in $^{17}$C we also couple three
channels for simplicity:
$$
|\frac{3}{2}^+\rangle=b_1|1d_{3/2}\otimes 0^+\rangle + b_2|1d_{3/2}\otimes
 2^+\rangle + b_3|2s_{1/2}\otimes 2^+\rangle.
$$

The system of coupled channel equation using potential expansion of
Eq.(\ref{taylor}) after calculation of angular matrix elements
$\langle ljI;JM|Y_{20}(\cos\theta')|l'j'I';JM\rangle$
 has the
following form:

\begin{eqnarray}
\nonumber
\left(T_i+V^{so}_i(r)+
V_0f(r)+
c_{ii}\beta v(r)+\epsilon_i-E\right)\chi_i(r)=\\
-\beta v(r)\sum_{i'\neq i}c_{ii'}\chi_{i'}(r),\,\,\,
\label{ccs_tay}
\end{eqnarray}
where index $i$ numbers the definite channel $|ljI\rangle$,
$V^{so}_i(r)$ is a spin orbit potential in channel $i$ and
$c_{ii'}$ is array of angular matrix elements (see Table
\ref{ame}).

\begin{table}
\caption{\label{ame}Angular matrix elements $c_{ii'}$ for the ground state
$\frac{1}{2}^+$ ($^{19}$C,$^{15}$C) and $\frac{3}{2}^+$
($^{17}$C)}
\begin{ruledtabular}
\begin{tabular}{cccc|cccc}
\,& $\frac{1}{2}^+$ &\,& \,&\,$\frac{3}{2}^+$ &\, &\,\\
\hline
$i$ & 1 & 2 & 3 & i & 1 & 2 & 3 \\
 1  & 0 & 0.218 & 0.178 & 1  & 0 & -0.126 & -0.126\\
 2  & 0.218 & 0.14 & 0.044 &  2  & -0.126 & 0 & 0.126 \\
 3  & 0.178 & 0.044 & 0.12 & 3  & 0.126 & 0.126 & 0 \\
\end{tabular}
\end{ruledtabular}
\end{table}

The solution of the system of coupled channel equations
Eq.(\ref{ccs_main}) or Eq.(\ref{ccs_tay}) gives the bound state
energies and the corresponding radial functions of the components
with their relative strengths. In the present calculation we used
the R-matrix method on Lagrange mesh (see Ref. \cite{Bay98} and
references therein) to solve the system of coupled channel
equations (\ref{ccs_main}).

\section{Coulomb breakup cross sections and dipole strength function}

In the semiclassical model of Coulomb excitation and breakup the process
of absorption of radiation  is treated quantum mechanically but the
projectile is assumed to move in a straight line (see Refs. \cite{Win79,Ber01}).
In the first order perturbation theory the process of Coulomb
excitation can be described as emission and absorption of virtual photons.
Thus, Coulomb dissociation spectrum is related to the dipole strength
function $dB(E1)/dE$ through
\begin{equation}
\frac{d\sigma_C}{dE}=\frac{16\pi^3}{9\hbar
  c}\frac{dB(E1)}{dE}
\frac{1}{E}\int_{b_0}^{\infty}2\pi bdb N_{E1}(E^*,b),
\label{sig_distr}
\end{equation}
where $E$ is the relative energy between the core and the neutron,
$E^*$ is an excitation energy,
$N_{E1}(\omega,b)$ is the number of virtual
photons, $\omega=E^*/\hbar$. $b_0$ is the cutoff parameter, which is
approximated by the sum of projectile and target radii. Here we assumed
sharp cutoff theory for Coulomb breakup.
Within the framework of a direct breakup mechanism, dipole strength
function is given by the transition matrix element
\begin{equation}
\frac{dB(E1)}{dE}=
\sum_M\left|<{\bf q}|\hat d|\Phi({\bf r})>\right|^2,
\,\,
\hat d=\frac{Ze}{A}rY_{1M},
\label{b_distr}
\end{equation}
where $<{\bf q}|$ represents the scattering state, $\Phi({\bf r})$ is the
ground state wave function of the nucleus and $\hat d$ is the electric
dipole operator.

The cluster model or {\it Yukawa+plane wave} approximation \cite{Bet35} is
usually used to calculate $dB(E1)/dE$ distributions.
In this approximation one assumes Yukawa
wave function for the ground state and a plane wave for a continuum
state and dipole strength function takes the form
\begin{equation}
\frac{dB(E1)}{dE}=SN_0\frac{3\hbar^2}
{\pi^2\mu}\left(\frac{Z_ce}{A_c+1}\right)^2\frac{\sqrt{E_B}E^{3/2}}
{(E+E_B)^4}
\label{y_distr}
\end{equation}
where $S$ is the spectroscopic factor for the $2s_{1/2}$ state and
$N_0$ is the normalization factor (see Refs. \cite{Ber91,Ber92,Bau92}).
$E_B$ is the
binding energy of the neutron in the ground state, $\mu$ is the
reduced mass of core+neutron and $A_c$ is the mass of the core.
It is seen that the the shape of the distribution depends only on the
binding energy $E_B$ and has a peak at $E=\frac{3}{5}E_B$.
The integral, $B(E1)$ is given by
\begin{equation}
B(E1)=\frac{3\hbar^2e^2}{16\pi E_B\mu}SN_0
\end{equation}
The validity of the {\it Yukawa+plane waves} approximation is based on
the fact that for low excitation energy $dB(E1)/dE$ is determined
by the outer part of the wave function.

When allowing the core to be excited (deformed or vibrational),
then the cross section can be expressed as the incoherent sum of
components $d\sigma (I)/dE$ corresponding to different
core states with spin $I$ populated after one
neutron removal. Furthermore for each core state the cross section
is further decomposed into an incoherent sum over contributions
from different angular momenta $j$ of the valence neutron in it's
initial state. Accordingly one has the general expression for a
final plane wave (n+core), $\langle {\bf q}|$,
\begin{eqnarray}
\nonumber
\frac{d\sigma (I)}{dE}\propto N_{E1}(E)
\sum_jC^2S(I,nlj)\times \\
\sum_m\left|\langle {\bf q}\otimes
I| \frac{Z_e}{A}rY_{1m}|\psi_{nlj}(r)\otimes
I\rangle\right|^2,
\label{cc_excl}
\end{eqnarray}
where the value of {\bf q} depends on the energy of the core state.

The above form for the exclusive Coulomb dissociation has an
analogous, one-neutron removal cross section (nuclear) obtained
within the eikonal approximation by Tostevin \cite{Tos99,Hus85} and used
extensively for spectroscopic study of radioactive nuclei.

When the inclusive Coulomb dissociation is required one, namely the
final state of the core is not observed, then the cross-section
becomes
\begin{equation}
\frac{d\sigma}{dE}=\sum_{I}\frac{d\sigma(I)}{dE}
\label{sigma_sum}
\end{equation}

The question we ask here is how large are the interference terms
present in the individual $d\sigma(I)/dE$ and in the
sum, Eq (\ref{sigma_sum})? These genuine quantal interference
terms arise naturally when calculating the matrix elements
$\langle{\bf q}|\frac{Z_c}{A}rY_{1m}|\psi_{nlj}\rangle$ using the
well known Bauer's expansion of the plain wave into partial waves
\begin{equation}
\langle{\bf q}|{\bf r}\rangle =
\frac{4\pi}{kr}\sum_{l,m}Y_{lm}(\hat q)Y_{lm}^*(\hat
r)(-i)^lj_l(kr).
\end{equation}

For each core state $I$ the dipole strength wave function is
calculated using the wave functions of Eq.(\ref{wf}) for the
ground states (with spin $J$) and plane waves for continuum states
(with spin $J'$) and has the following form \cite{Nun95}:

\begin{equation}
\frac{dB(E1,I,J\rightarrow J')}{dE}=\frac{\hat
J^2}{\hat{J'^2}}\left| \sum_{l,j,l',j'}{\cal M}_{lj,l'j'}^I{\cal
R}_{lj,l'j'}^I \right|^2, \label{pcdbde}
\end{equation}
where angular part of the matrix element is given by:
\begin{eqnarray}
\nonumber
 {\cal M}_{lj,l'j'}^I= \frac{Z_ce}{A_c+1}\sqrt{\frac{3}{4\pi}}
\hat J\hat j\hat j'\times\\
(-)^{J+I+j+j'-\frac{1}{2}}
 \left\{ \begin{array}{ccc}
\textstyle J'& \textstyle J& \textstyle 1\\
\textstyle j & \textstyle j' & \textstyle I \\
\end{array} \right\}
\left( \begin{array}{ccc}
\textstyle j'& \textstyle j& \textstyle 1\\
\textstyle \frac{1}{2} & \textstyle -\frac{1}{2} & \textstyle 0 \\
\end{array} \right)
\end{eqnarray}
and the radial part of the matrix element:
\begin{equation}
 {\cal R}_{lj,l'j'}^I=\sqrt{\frac{2\mu k}{\hbar^2\pi}}
 \int_{0}^{\infty}\chi_{ljI}^J(r)rj_{l'}(kr)dr,
\end{equation}
where $\chi_{ljI}^J(r)$ are g.s. radial wave functions of
Eq.(\ref{wf}) and $j_{l'}(kr)$ are the spherical Bessel functions.

The expression for the dipole strength function of
Eq.(\ref{pcdbde}) contains the coherent sum of contributions from
different channels (but with the same core state), which appear
because of the expansion of the total wave function of the ground
state in terms of the core states. This gives
rise to the interference terms in the dipole strength and
consequently in the cross section.

To compare with experimental data, one needs to sum the expression
(\ref{pcdbde}) over allowed final angular momentum $J'$. The spin
of the ground state of  $^{19}$C and  $^{15}$C is
$J=\frac{1}{2}^+$ and for the final state we have
$J'=\{\frac{1}{2}^-,\frac{3}{2}^-\}$. For the g.s. of $^{17}$C
with $J=\frac{3}{2}^+$ one also has
$J'=\{\frac{1}{2}^-,\frac{3}{2}^-\}$.

The contributions of different terms entering Eq.(\ref{pcdbde})
will be analyzed in the next Section.

\section{Application}
\begin{table}
\caption{\label{table2}Parameters of the model used to describe
$^{15}$C,$^{17}$C
  and $^{19}$C and the corresponding structure of the ground state
  wave functions. The values of the radius $R_0$, diffuseness $a$
  , spin-orbit depth $V^{so}_0$ and deformation parameter $\beta$ 
  are the same as in Ref. \cite{Rid98}.}
\begin{ruledtabular}
\begin{tabular}{cccc}
parameter & $^{19}$C & $^{17}$C & $^{15}$C\\
\hline
{\bf potential} & & & \\
$R_0$ (fm) &  3.00 &2.82 & 2.45\\
$a$ (fm)   & 0.65& 0.65& 0.65 \\
$V^{so}_0$ (MeV) & 6.5&6.5 &6.5 \\
$V^{ws}_0$ (MeV) & 42.95& 59.33& 71.12\\
$\beta$    & 0.5 & 0.55 & 0.42 \\
$\epsilon_{2^+}$ (MeV) & 1.2 & 1.776 & 7.1 \\
$S_{n}$ (MeV) & 0.65 & 0.73& 1.22\\
\hline
{\bf results} & & & \\
g.s. structure & $1/2^+$ & $3/2^+$ & $1/2^+$ \\
(\%)  & 71.5 $s_{1/2}\otimes0^+$&
  14.5  $d_{3/2}\otimes0^+$  &
 87.2 $s_{1/2}\otimes0^+$\\
                      & 25.3 $d_{5/2}\otimes2^+$&
  76.5 $s_{1/2}\otimes2^+$                            &
10.9 $d_{5/2}\otimes2^+$\\
                      & 3.2 $d_{3/2}\otimes2^+$&
                        9.0  $d_{3/2}\otimes2^+$     &
1.9 $d_{3/2}\otimes2^+$\\
\end{tabular}
\end{ruledtabular}
\end{table}
In this section we present our calculations for the Coulomb
dissociation of the carbon isotopes $^{19}$C, $^{17}$C and $^{15}$C
in the field of $^{208}$Pb.
The parameters of particle-core model used here for each of the carbon
isotopes are given
in Table \ref{table2}.

\subsection{$^{19}$C}

\begin{figure*}
\begin{tabular}{ll}
\includegraphics[scale=0.5]{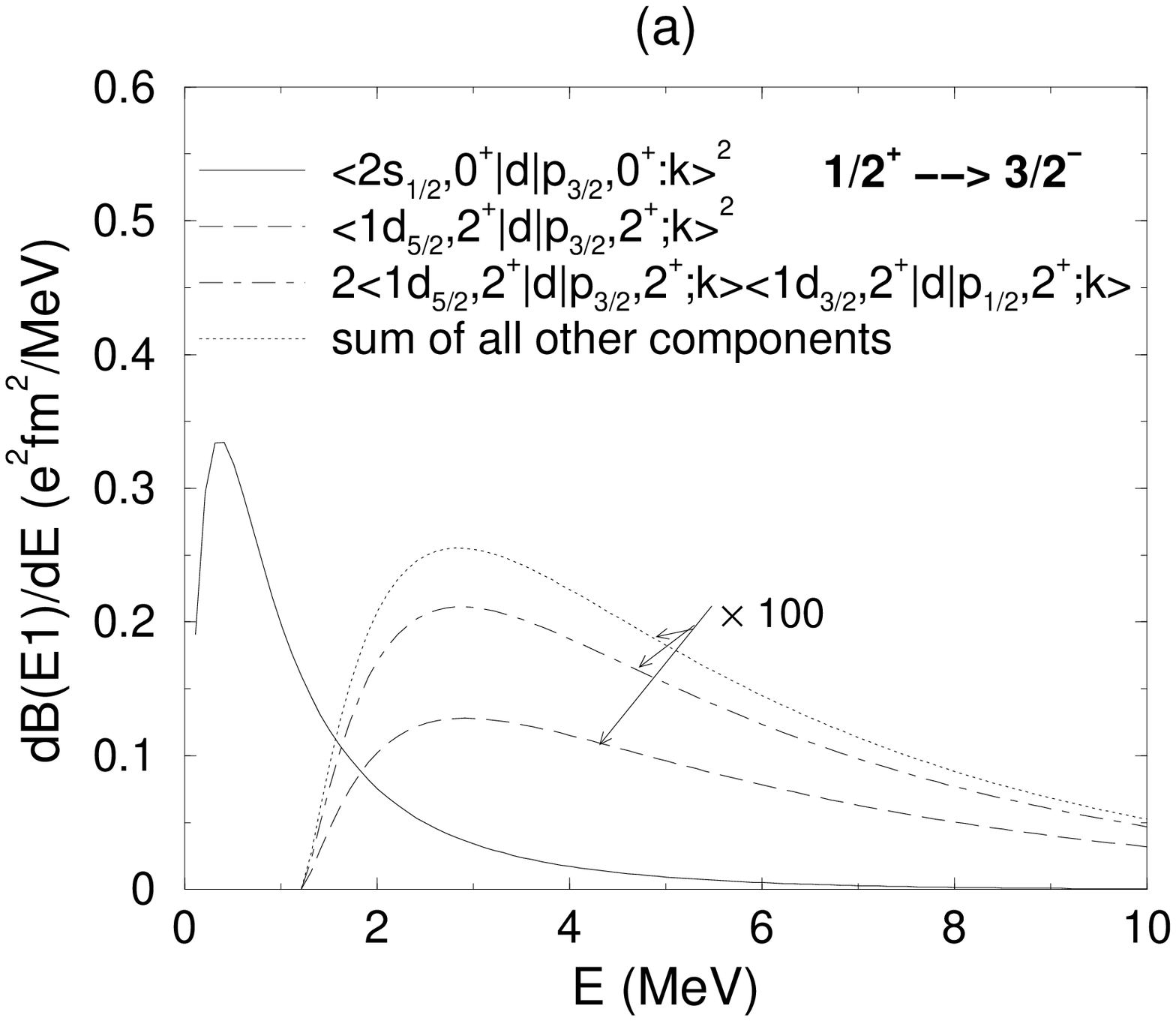} &
\includegraphics[scale=0.5]{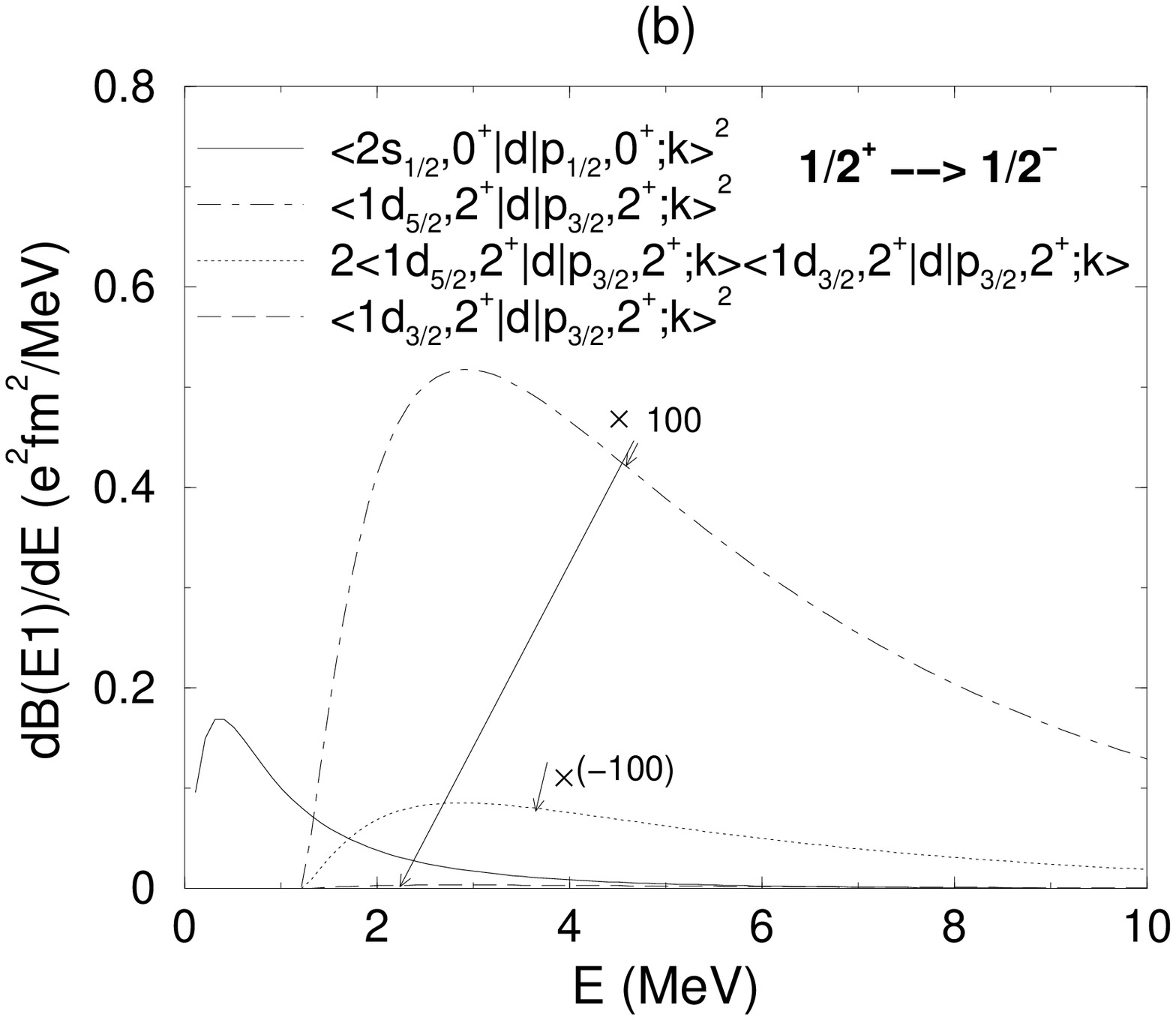}\\
\includegraphics[scale=0.5]{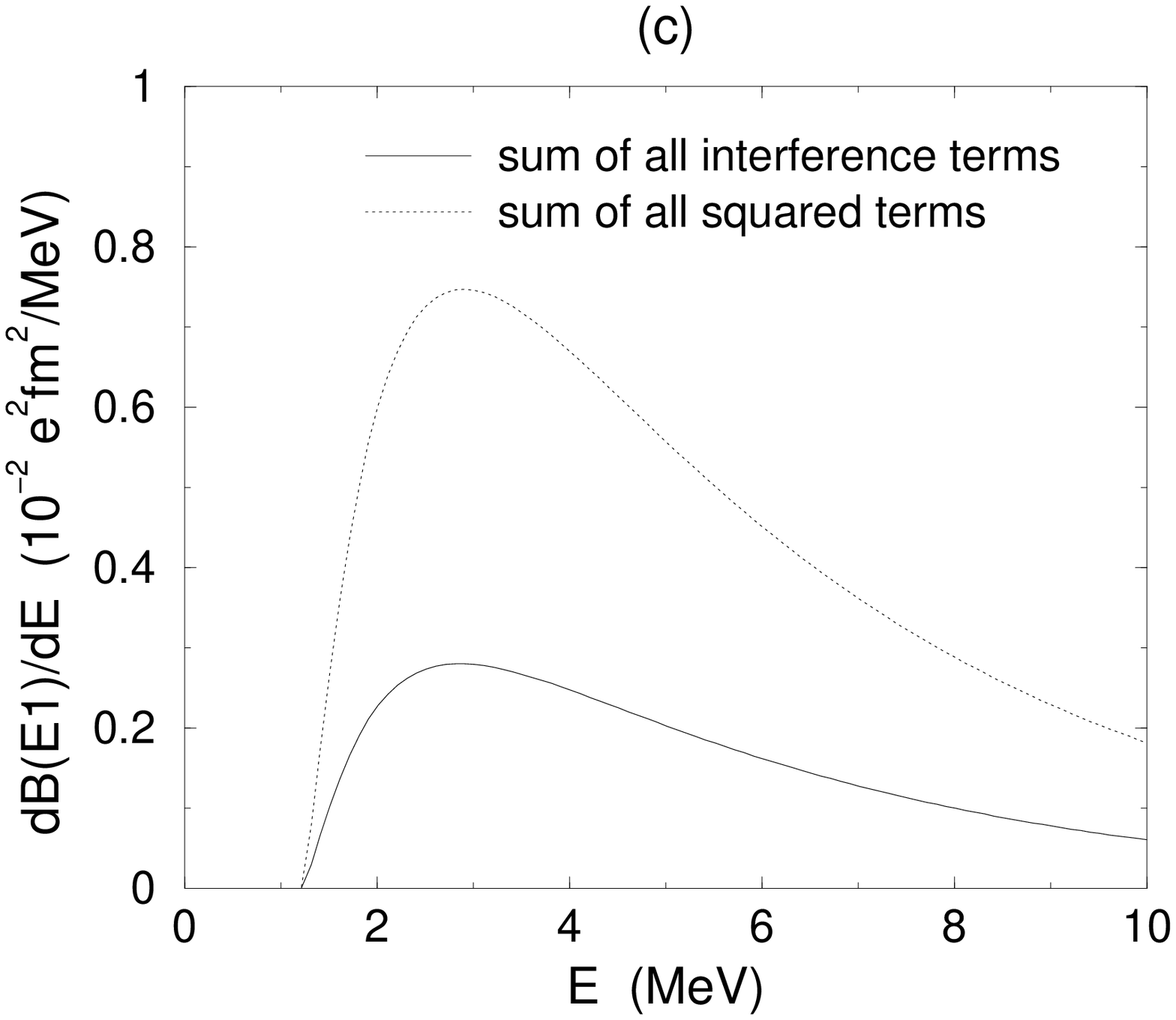} &
\includegraphics[scale=0.5]{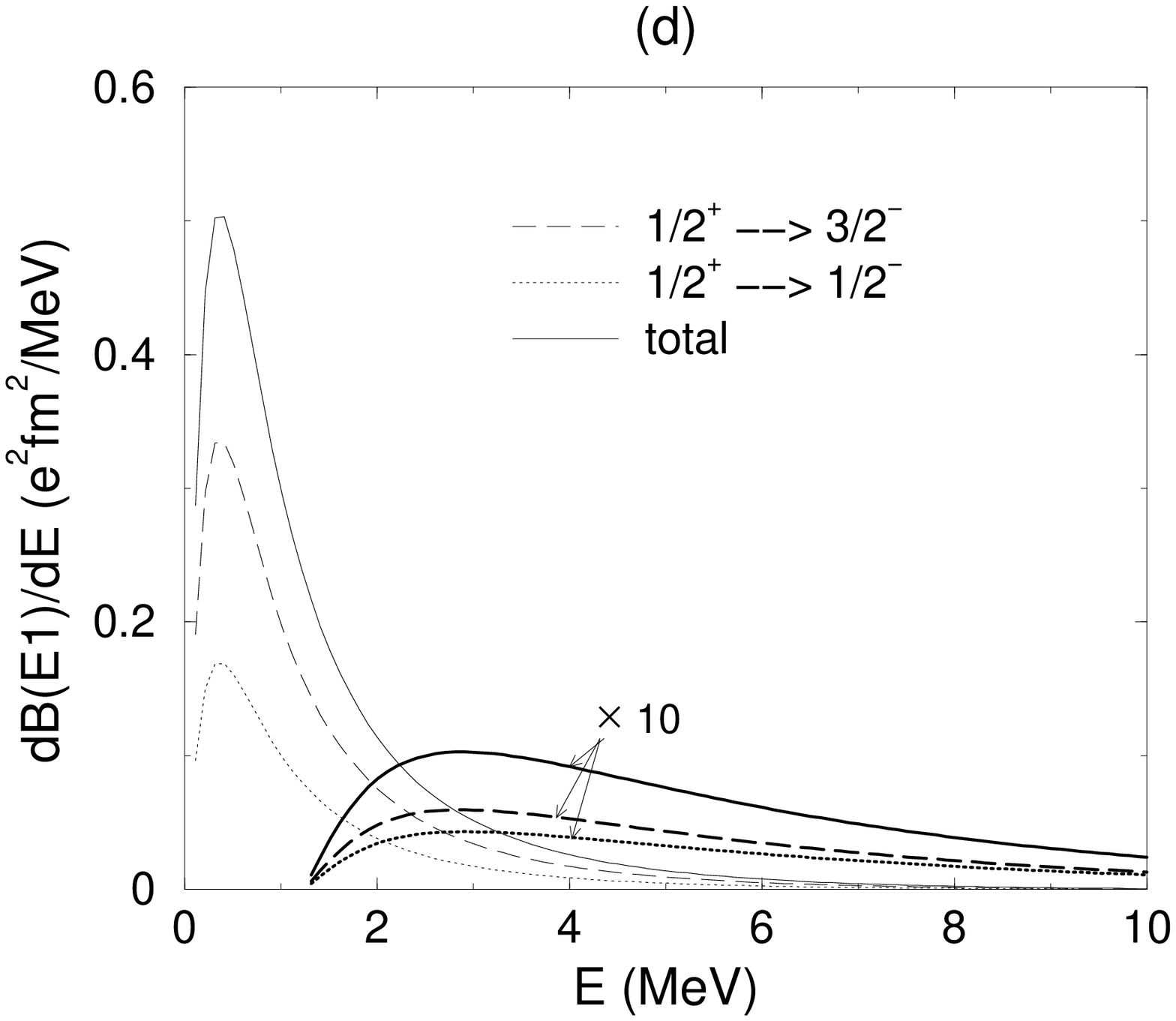}\\
\end{tabular}
\caption{Contributions to the dipole strength function for $^{19}$C.
(a)Dominant contributions to the dipole strength for
$\frac{1}{2}^+\longrightarrow\frac{3}{2}^-$ transition;(b)All
contributions to the dipole strength for $\frac{3}{2}^+\longrightarrow\frac{1}{2}^-$;
(c)Comparison of overall contribution of squared terms(contribution of
$0^+$ is not shown here) and interference terms; (d)overall comparison
of contribution of $\frac{1}{2}^+\longrightarrow\frac{3}{2}^-$ and
and $\frac{3}{2}^+\longrightarrow\frac{1}{2}^-$ and their sum for
$0^+$ contributions (thin lines)and $2^+$ contributions(thick lines);}
\label{Fig1}
\end{figure*}

The role of the transitions to the core excited states in the
Coulomb breakup of $^{19}$C (and $^{11}$Be) on $^{208}$Pb was
analyzed before in Ref. \cite{Shy01}, where the integrated partial
cross sections and momentum distributions for the ground state as
well as excited bound states of core nuclei were calculated within
the finite-range DWBA as well as within the adiabatic model of the
Coulomb breakup. It was found that the transitions to excited
states of the core are quite weak and the interference effects are
suppressed.

In our calculation, there are 7 coherent contributions to the
$\frac{1}{2}^+\longrightarrow \frac{3}{2}^-$ transition and 4
coherent contributions to the
$\frac{1}{2}^+\longrightarrow\frac{1}{2}^-$ transition. In fact,
in the $\frac{1}{2}^+\longrightarrow\frac{3}{2}^-$ case, we found
out that the dominant contribution is the direct
$s_{1/2}\longrightarrow p_{3/2}$ one (see Fig.\ref{Fig1}a). All
the other 6 contributions add up to a $6\%$ of the total,
confirming the result of Ref. \cite{Shy01}. The contributions of
core excited state $2^+$ are very small and in order to make them
visible they are multiplied by 100 in Fig.\ref{Fig1}a and
Fig.\ref{Fig1}b and by 10 in Fig.\ref{Fig1}c.
 In the
$\frac{1}{2}^+\longrightarrow\frac{1}{2}^-$ transition, besides
the dominant $s_{1/2}\longrightarrow p_{1/2}$ one, there is only
one interference term which add up destructively, namely $2\langle
1d_{5/2}, 2^+|\hat d|p_{3/2},2^+;k\rangle \langle 1d_{3/2},
2^+|\hat d|p_{3/2},2^+;k\rangle$. The other 2 contributions, add
up to about few percents, as Fig.\ref{Fig1}b shows. Total
contributions of all squared terms (core excited components only)
and interference terms are compared in Fig.\ref{Fig1}c.

The total
$\frac{1}{2}^+\longrightarrow\frac{3}{2}^-$
and $\frac{1}{2}^+\longrightarrow\frac{1}{2}^-$ transitions are shown
in Fig.\ref{Fig1}d together with their sum, to be compared with the data.

\begin{figure}
\centerline{\includegraphics[scale=0.5]{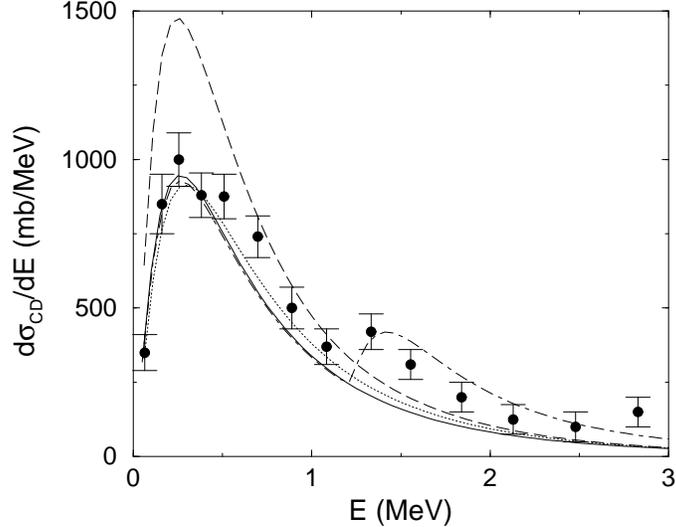}}
\caption{$^{19}$C+$^{208}$Pb at 67A MeV Coulomb breakup cross sections
  calculated in different models. Solid line stands for the cluster
  model calculation with $S$=0.67; dashed line -- present calculation
  with $\epsilon_{2^+}$=1.62 MeV; dotted line -- present calculation with
  $\epsilon_{2^+}$=1.2 MeV; dot-dashed line -- toy model calculation (see text).
Data from \cite{Nak99}.}
\label{19c_cc}
\end{figure}

When compared to the data, the major low-energy peak is quite
nicely reproduced, with 70$\%$ $s$-contribution and 30$\%$
$d$-contribution. The data also exhibit some structure at higher
excitation energies, which can not be accounted for by the
particle-rotor model. In fact, we have already taken the liberty
of reducing the energy of the $2^+$ state of the core (since the
rotor is found inside the halo nucleus and changes may ensue) to
see whether this, with accompanying change in the relative energy
of the halo neutron, can reproduce better the data. We found out
that the combination $\epsilon_{2^+}=1.2$ MeV and $S_n=0.65$ MeV
gives the best account, but still misses the strength at higher
energies. Since the total contribution to the cross section
involves the incoherent sum of the
$\frac{1}{2}^+\longrightarrow\frac{3}{2}^-$ and
$\frac{1}{2}^+\longrightarrow\frac{1}{2}^-$, and the former
exhibits a significant structure besides the main peak, we were
tempted to consider a two-cluster model for the cross-section, one
peaks at $E=\frac{3}{5}E_B$, while the other at
$E=\frac{3}{5}(X-E_B)$, with X related to the excitation energy of
the $2^+$ state in the deformed core. The result of our
calculation is shown in Fig.\ref{19c_cc}. The ``spectroscopic
factors" attached to the two cluster response are, respectively,
0.7 and 0.3. The above toy model is instructive, as it points to
ways of improving the particle-excited core model. Of course, the
Coulomb dissociation, being, what is known as, elastic breakup (no
target excitation) may have to be corrected owing to possible
contributions of nuclear-Coulomb interference \cite{Typ01}
 and inelastic breakup.

\subsection{$^{17}$C}

In the dissociation of $^{17}$C, whose ground state spin is
$\frac{3}{2}^+$, we again sum incoherently two transitions
$\frac{3}{2}^+\longrightarrow\frac{3}{2}^-$ and
$\frac{3}{2}^+\longrightarrow\frac{1}{2}^-$.
Unlike the case of breakup of $^{19}$C, here we have $^{17}$C
mainly with the core excited contribution in the ground state
wave function (see Table \ref{table2}). Therefore terms
$d_{3/2}\longrightarrow p_{3/2}$ and $d_{3/2}\longrightarrow p_{1/2}$,
which corresponds to the contribution of the $0^+$ in the core,
are now small.

\begin{figure*}
\begin{tabular}{ll}
\includegraphics[scale=0.5]{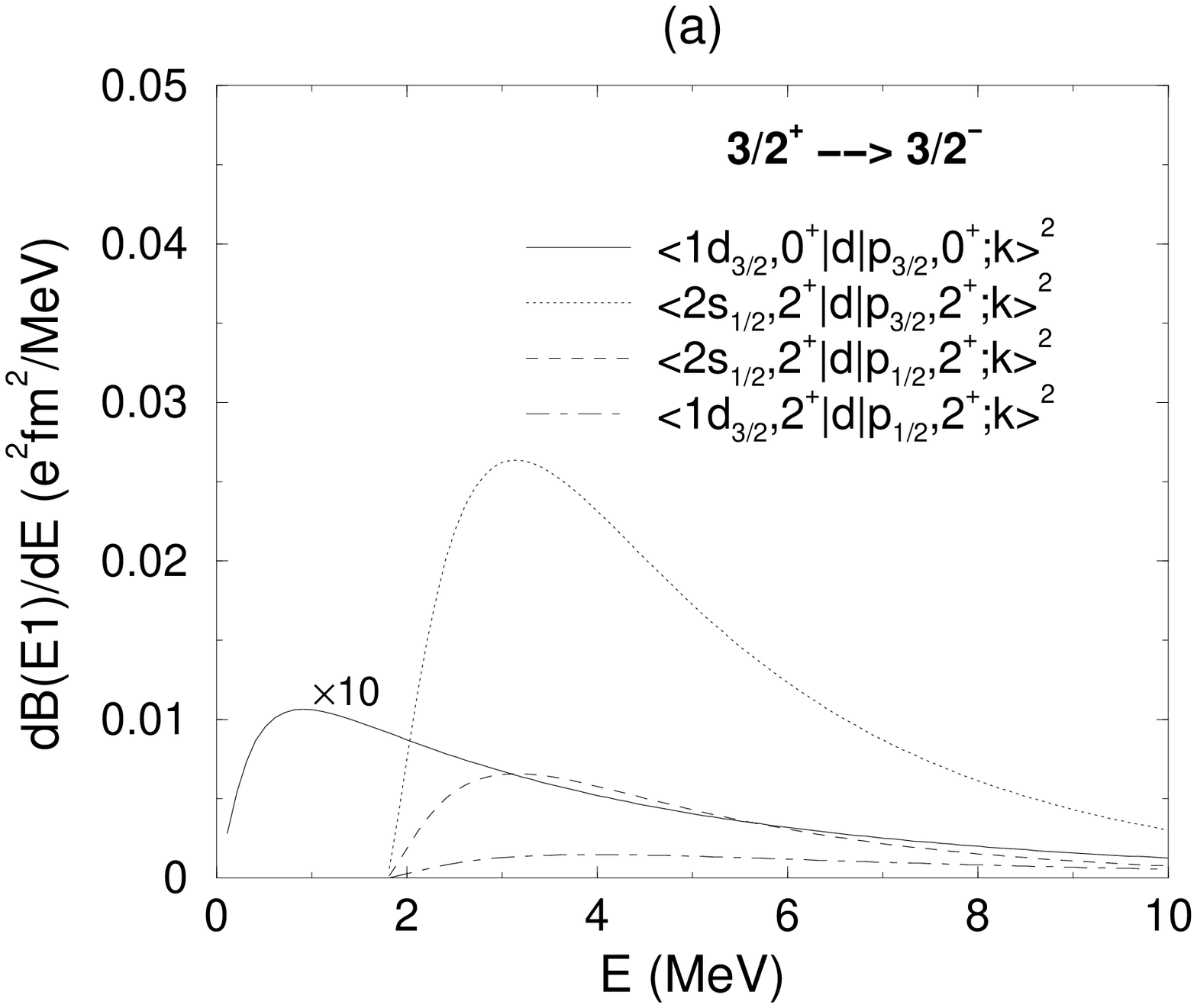} &
\includegraphics[scale=0.5]{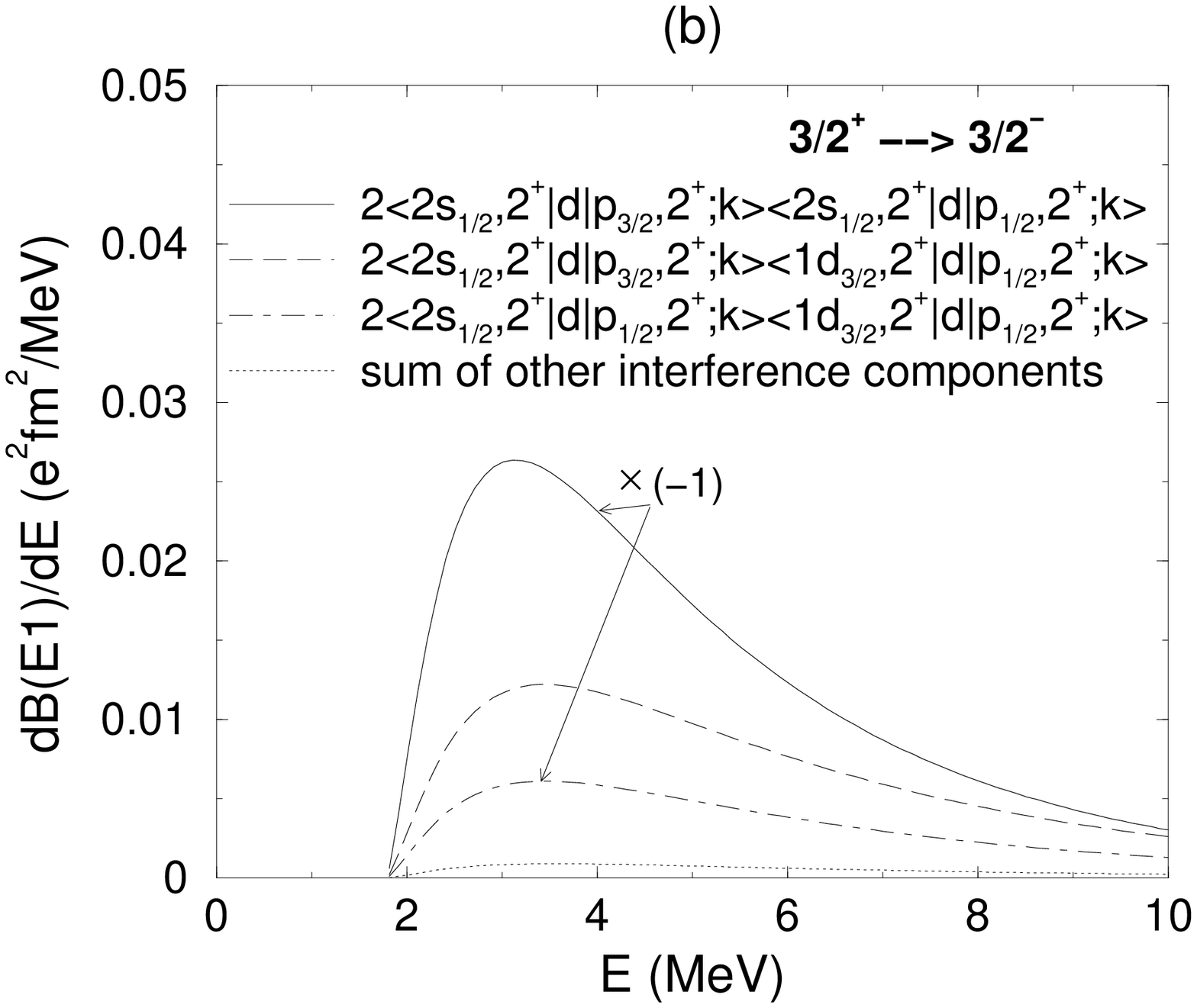}\\
\includegraphics[scale=0.5]{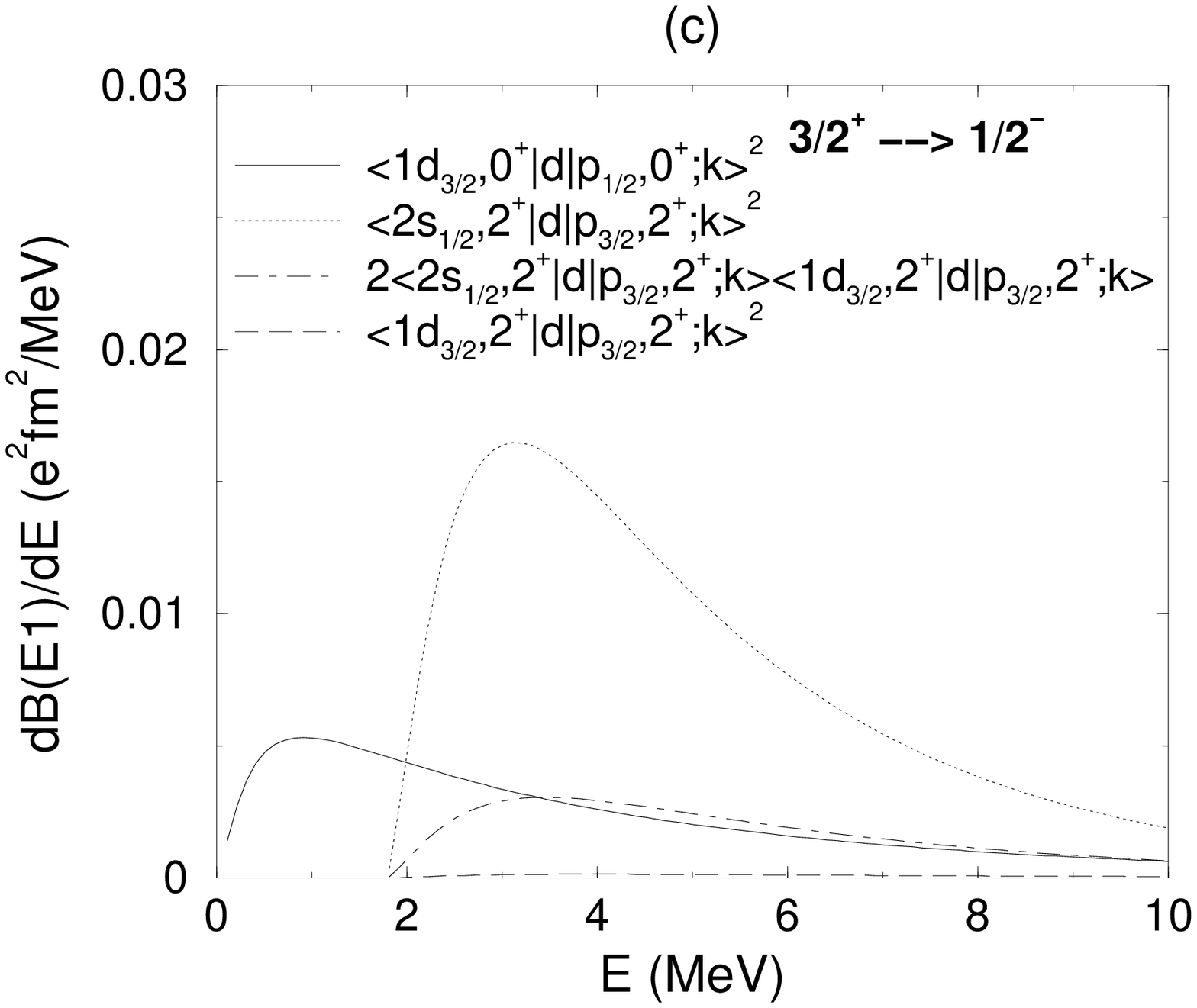} &
\includegraphics[scale=0.5]{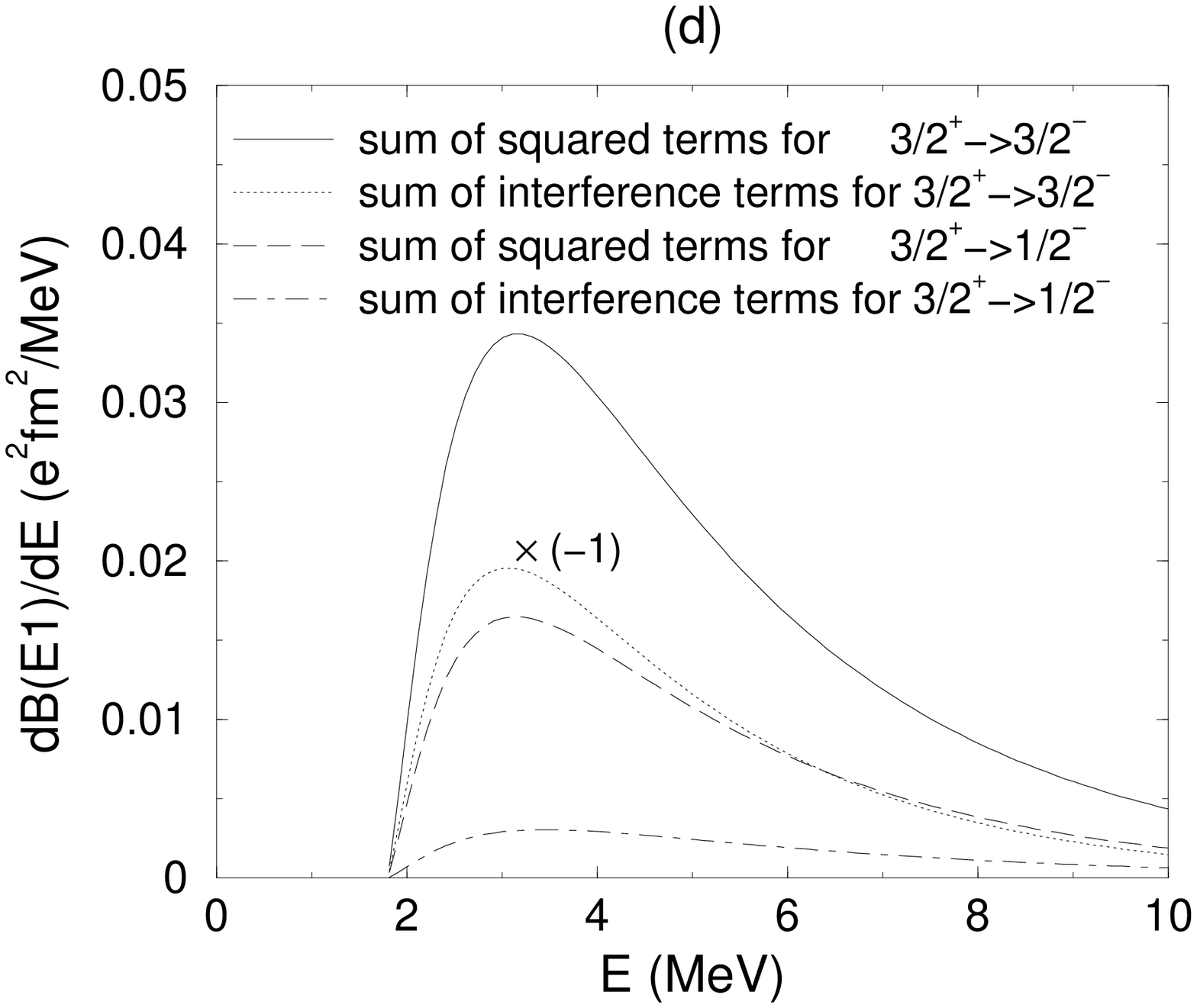}\\
\end{tabular}
\caption{Contributions to the dipole strength function for $^{17}$C.
(a)All squared contributions to the dipole strength for
$\frac{3}{2}^+\longrightarrow\frac{3}{2}^-$;
(b)Dominant interference contributions for
$\frac{3}{2}^+\longrightarrow\frac{3}{2}^-$;
(c)All contributions to the dipole strength for $\frac{3}{2}^+\longrightarrow\frac{1}{2}^-$;
(d)Comparison of overall contribution of squared terms (contribution
of $0^+$ is not shown here)and interference terms.
}
\label{Fig3}
\end{figure*}

For $\frac{3}{2}^+\longrightarrow\frac{3}{2}^-$
there are all together 10 contributions of which
6 are interference terms (see Fig.\ref{Fig3}a).
For $\frac{3}{2}^+\longrightarrow\frac{3}{2}^-$ transition,
the dominant terms $\frac{3}{2}^+\longrightarrow\frac{1}{2}^-$
and
$2\langle 2s_{1/2}, 2^+|\hat d|p_{3/2},2^+;k\rangle\langle 1d_{3/2}, 2^+|\hat d|p_{1/2},2^+;k\rangle$
cancel each other.
As Fig.\ref{Fig3} shows, the overall strength is
predominantly due to the $2^+$ state in the core and is rather
small because of the cancellation of the two dominant terms above.
The contribution of the $0^+$ in the core,
namely $|\langle 1d_{3/2}, 0^+|\hat d|p_{3/2},0^+;k\rangle|^2$ is
but a few percent.
This shows clearly that in the
$\frac{3}{2}^+\longrightarrow\frac{3}{2}^-$ transition quantum
interference and core excitation are very important.

The transition $\frac{3}{2}^+\longrightarrow\frac{1}{2}^-$
contains altogether three squared and one interference terms,
the dominant squared one is $|\langle 2s_{1/2}, 2^+|\hat
d|p_{3/2},2^+;k\rangle|^2$ and the only one interference term
$2\langle 2s_{1/2}, 2^+|\hat d|p_{3/2},2^+;k\rangle\langle 1d_{3/2}, 2^+|\hat d|p_{3/2},2^+;k\rangle$
positive and
almost equal the direct term $d_{3/2}\longrightarrow p_{1/2}$.

Transitions
$\frac{3}{2}^+\longrightarrow\frac{1}{2}^-$ and
$\frac{3}{2}^+\longrightarrow\frac{3}{2}^-$
contribute almost equally into total.
If one considers the full contribution
of the interference terms, namely from the
$\frac{3}{2}^+\longrightarrow\frac{3}{2}^-$ and
$\frac{3}{2}^+\longrightarrow\frac{1}{2}^-$ transitions, one get
just about 30$\%$ of the total squared terms.
Note, that in total interference terms contribute destructively.
To within error bar
accuracy, one may conclude that the $dB(E1)/dE$ distribution for
$^{17}$C is composed of an incoherent sum of several squared terms
that peak at about the same energy, corresponding, to the
$2s_{1/2}\otimes 2^+$ component of the g.s. configuration. This
attests to the validity of the simple cluster model even in this
subtle quantal  system.

\begin{figure}
\centerline{\includegraphics[scale=0.5]{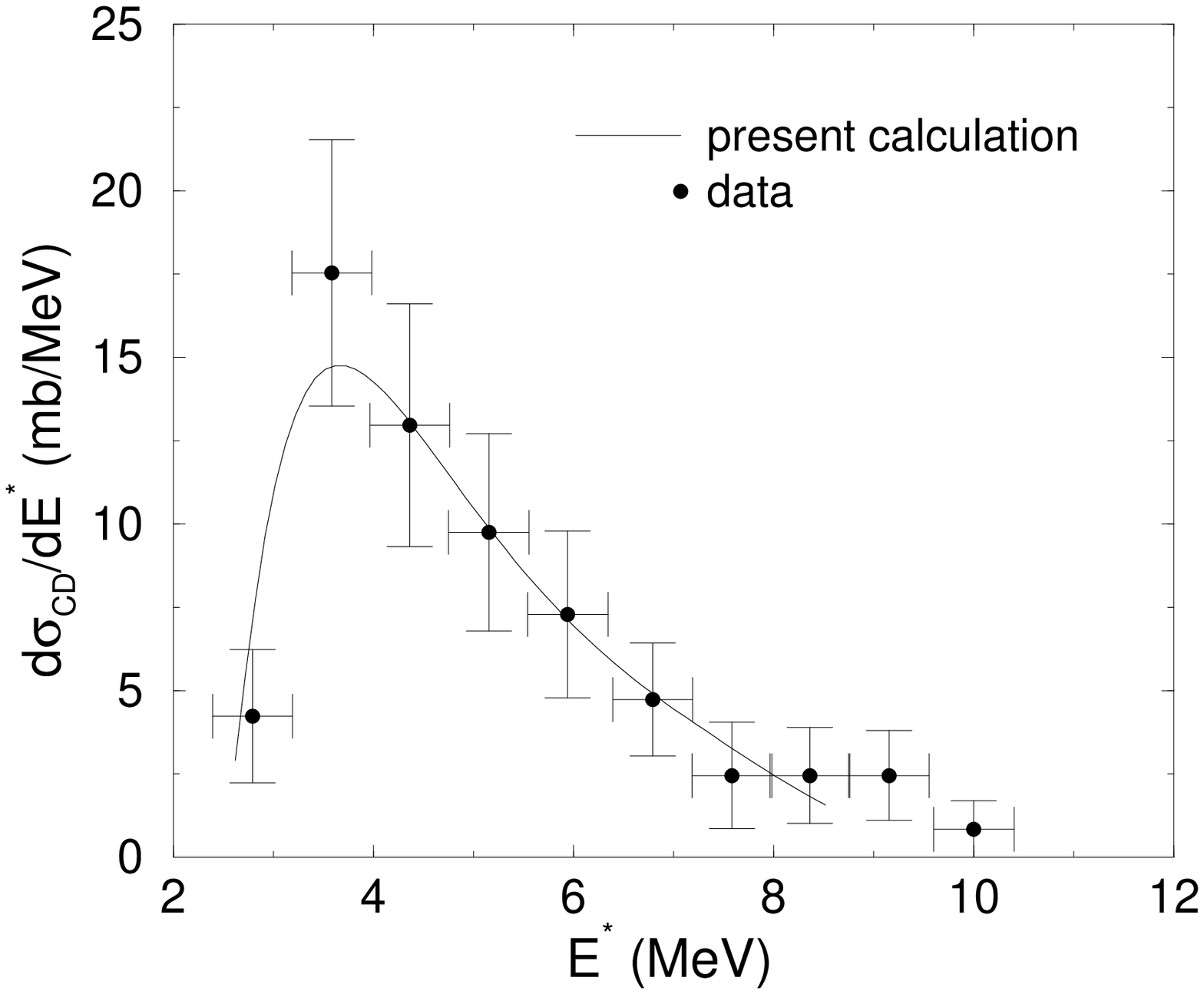}}
\caption{$^{17}$C+$^{208}$Pb at 495A MeV Coulomb breakup cross sections
  calculated in different models. Data from \cite{Pra03}. $E^*$ is an
  excitation energy.
(Here the theoretical cross sections were convoluted with the
experimental response functions.)}
\label{17c_cc}
\end{figure}

In Ref. \cite{Pra03}, the final channel was clearly identified as
$^{16}$C($2^+$)+n. Therefore, to be correct, one has to compare
the data with the exclusive Coulomb dissociation cross-section,
Eq.(\ref{cc_excl}). However, as we have seen, the core $0^+$ state contribution
is at most 10$\%$ of the peak value of the $2^+$ contribution.
Further, it is mostly concentrated in the relative energy range
$0<E<2$ MeV. Thus one may safely calculate the inclusive
cross-section and ignore the low energy part. The comparison with
the data is shown in Fig.\ref{17c_cc}. It can be seen, that
the present calculation reproduces the data.

\subsection{$^{15}$C}

\begin{figure}
\centerline{\includegraphics[scale=0.5]{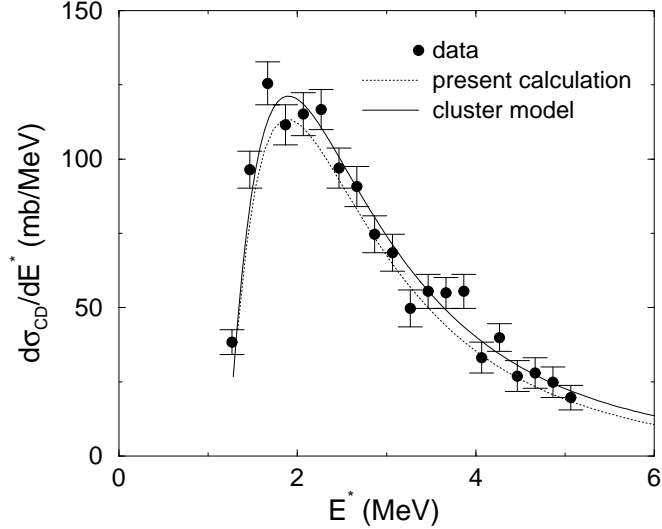}}
\caption{$^{15}$C+$^{208}$Pb at 605A MeV Coulomb breakup cross sections
  calculated in different models. Data from \cite{Pra03}. $E^*$ is an
  excitation energy.
(Here the theoretical cross sections were convoluted with the
experimental response functions.)
}
\label{15c_cc}
\end{figure}

The last example is the calculation of the Coulomb breakup of
$^{15}$C. 
Strictly speaking, we cannot use the rotor model for the core, 
as it was experimentally shown, that the first
excited state of the core is not $2^+$ but $1^-$ at around 7 MeV \cite{Ajz91}.
But, as it was shown in Ref. \cite{Rid98}, particle-rotor model
reproduces the spectroscopic number rather well. This calculation
gives 87$\%$ of the core inert state. Because of this the
contribution of core excited components is very small and we do
not include it here, that is we only include contributions
$2s_{1/2}\longrightarrow p_{3/2}$ and $d_{3/2}\longrightarrow
p_{1/2}$ to the dipole strength function. The resulting cross
section is shown in Fig.\ref{15c_cc} along with the result of the
cluster model Eq.(\ref{y_distr}) . Again the simple cluster
picture works nicely for this system.

\section{Conclusions}

We have used the particle-deformed core model wave functions to
calculate the Coulomb dissociation of $^{15}$C, $^{17}$C and
$^{19}$C in the semiclassical model. We have shown that whereas
interference effects are important in the non-halo isotope
$^{17}$C, they are practically insignificant in $^{15}$C and
$^{19}$C. This fact points to the validity of the cluster picture
in the latter cases where it emphasizes the subtle quantal nature
of the former, mostly core-excited projectile.

\acknowledgements
Authors would like to acknowledge the partial support of Brazilian
agencies FAPESP and CNPq.

\end{document}